%% file: main.tex
\documentclass[conference]{IEEEtran}

\usepackage{cite}
\usepackage{amsmath,amssymb,amsfonts}
\usepackage[ruled,linesnumbered]{algorithm2e}
\usepackage{graphicx}
\usepackage{textcomp}
\usepackage{xcolor}
\usepackage{booktabs}

\usepackage[hidelinks]{hyperref} 
\usepackage{orcidlink}
\usepackage[caption=false, font=footnotesize]{subfig}

\usepackage{makecell}
\usepackage[all]{hypcap} 
\usepackage{siunitx} 
\usepackage[nonumberlist]{glossaries}
\usepackage{glossary-inline}
\usepackage[capitalize]{cleveref} 

\loadglsentries[main]{glosfile}
\glsdisablehyper 

\usepackage[colorinlistoftodos,prependcaption,textsize=tiny]{todonotes} 

\allowdisplaybreaks

\newcommand{\tool}{SCIM MILQ}

\makeatletter
\newcommand{\linebreakand}{%
  \end{@IEEEauthorhalign}
  \hfill\mbox{}\par
  \mbox{}\hfill\begin{@IEEEauthorhalign}
}
\makeatother

\newcommand\copyrighttext{%
  \footnotesize  © 20xx IEEE. Personal use of this material is permitted. Permission from IEEE must be
obtained for all other uses, in any current or future media, including
reprinting/republishing this material for advertising or promotional purposes, creating new
collective works, for resale or redistribution to servers or lists, or reuse of any copyrighted
component of this work in other works.}
\newcommand\copyrightnotice{%
\begin{tikzpicture}[remember picture,overlay]
\node[anchor=south,yshift=10pt] at (current page.south) {\fbox{\parbox{\dimexpr\textwidth-\fboxsep-\fboxrule\relax}{\copyrighttext}}};
\end{tikzpicture}%
}

\begin{document}

\title{\tool{}: An HPC Quantum Scheduler  \\
}

\author{\IEEEauthorblockN{Philipp Seitz\IEEEauthorrefmark{1}\orcidlink{0000-0003-3856-4090}, Manuel Geiger\IEEEauthorrefmark{1}\orcidlink{0000-0003-3514-8657}, Christian Ufrecht\IEEEauthorrefmark{2}\orcidlink{0000-0003-4314-9609}, \\
Axel Plinge\IEEEauthorrefmark{2}\orcidlink{0000-0001-7757-2953}, Christopher Mutschler\IEEEauthorrefmark{2}\orcidlink{0000-0001-8108-0230}, Daniel~D.~Scherer\IEEEauthorrefmark{2}\orcidlink{0000-0003-0355-4140} and Christian~B.~Mendl\IEEEauthorrefmark{1}\IEEEauthorrefmark{3}\orcidlink{0000-0002-6386-0230}}
\IEEEauthorblockA{\IEEEauthorrefmark{1}\emph{Technical University of Munich}\\
\emph{TUM School of Computation, Information and Technology}\\
\emph{Department of Computer Science}\\
Boltzmannstra{\ss}e 3, 85748 Garching, Germany \\
\{philipp.seitz, manuel.geiger, christian.mendl\}@tum.de
}
\IEEEauthorblockA{\IEEEauthorrefmark{2}\emph{Fraunhofer IIS}\\
\emph{Fraunhofer Institute for Integrated Circuits IIS} \\
\emph{Division Positioning and Networks}\\
Nordostpark 84, 90411 Nürnberg, Germany}
\{christian.ufrecht, axel.plinge, christopher.mutschler, daniel.scherer2\}@iis.fraunhofer.de
\IEEEauthorblockA{\IEEEauthorrefmark{3}\emph{Technical University of Munich}\\
\emph{TUM Institute for Advanced Study}}
}

\maketitle

\begin{abstract}
With the increasing sophistication and capability of quantum hardware, its integration, and employment in \gls{HPC} infrastructure becomes relevant.
This opens largely unexplored access models and scheduling questions in such quantum-classical computing environments, going beyond the current cloud access model.
\tool{} is a scheduler for quantum tasks in \gls{HPC} infrastructure.
It combines well-established scheduling techniques with methods unique to quantum computing, such as circuit cutting.
\tool{} can schedule tasks while minimizing the makespan, i.e., the time that elapses from the start of work to the end, improving on average by \SI{25}{\percent}.
\emph{Additionally}, it reduces the noise in the circuit by up to \SI{10}{\percent}, increasing the outcome's reliability.
We compare it against an existing baseline and show its viability in an \gls{HPC} environment.
\end{abstract}

\copyrightnotice{}

\glsresetall

\section{Introduction}\label{sec:Introduction}

Quantum computing has gained traction in the field of \gls{HPC}, starting with integrating quantum hardware into compute centers.
The prospect of quantum advantage is especially promising for application in the \gls{HPC} domain as large-scale problems profit most from the potential speedup.
Driven by the progress of quantum hardware, the quantum computing field moves from the so-called \gls{NISQ}\cite{Preskill2018} era toward quantum computing at the utility scale.
While error correction and fault-tolerant quantum computation rapidly progress as ~\cite{Bluvstein2024}, noise will remain a limiting factor in near-term quantum hardware.
Until full error correction is available, most algorithms that promise exponential quantum advantage, such as Shor's algorithm~\cite{shor1994algorithms}, are infeasible.

A new noisy but large-scale category emerges between \gls{NISQ} and fault-tolerant quantum computing.
In this category, quantum devices offer sufficient qubit numbers for real-world applications.
Interlinked devices, as outlined in the IBM roadmap~\cite{ibm2022}, are on the verge of becoming available.
These hardware improvements offer new possibilities, which need to be provided to users, specifically in \gls{HPC}.
This requires novel approaches on the software side to keep up with the pace of hardware capabilities.
W
This paper proposes \tool{} (\textbf{S}uper\textbf{C}omputing \textbf{I}ntegrated \textbf{M}ethods for MILQ), a runtime component for hybrid \gls{HPCQC} applications.
It includes classical scheduling techniques adapted to the quantum domain and quantum circuit cutting incorporated at runtime in two novel scheduling approaches for quantum circuits.
We built \tool{} on top of our previous work \emph{MILQ}~\cite{seitz2023}, addressing significant performance weaknesses that would render it impractical for specific scenarios, which will be discussed later.
We notably introduce a reinforcement learning agent with circuit-cutting capabilities to consider noise when scheduling.
With \tool{}, we provide a crucial part of the runtime environment, which can be tailored to the specific needs of compute centers.
The implementation is available as a repository on GitHub at \url{https://github.com/qc-tum/milq}.

\section{Motivation}\label{sec:Motivation}

Currently, most vendors in the quantum computing field offer access to their hardware via a cloud service.
When users want to run their algorithms, they explicitly choose a device and are enqueued in a job queue specific to that device.
While this was sufficient for the early days of quantum computing with small algorithms and small devices, there are now more users, bigger devices, and more complicated algorithms, which leads to long wait times.
Two problems are prominent.
On the one hand, even small circuits use the whole device exclusively, leading to underutilization of the hardware.
On the other hand, some algorithms require a lot of qubits and might require the largest device, leading to an unbalanced distribution of jobs.
By combining scheduling with circuit cutting, both problems can be solved.
Multiple small circuits can be scheduled to run on the same device in parallel, possibly subjected to noise restrictions.
Large circuits can be cut such that the subcircuits can be scheduled on smaller, less busy devices.
In total, this increases throughput and decreases wait times for users.

With \emph{MILQ}~\cite{seitz2023}, we provided a prototype to solve this problem.
It automatically cuts, combines, schedules, and reconstructs circuits.
Still, it remains impractical in \gls{HPC} environments due to the generally higher load.
Solving the scheduling problem exactly is only tractable for small problem instances.
With increasing circuit sizes, the number of associated jobs after cutting increases exponentially, rendering the problem infeasible.
This can be adjusted by improving two components.
Involving the cutting decisions in the scheduler, which improves upon the overall number of circuits, and more capable scheduling algorithms, which can take circuit cutting into account. 

Additionally, \emph{MILQ} lacks the techniques shared by classical schedulers, for example, a uniform job submission system, preventing easy adoption.

\section{Background and Related Work}\label{sec:BackgroundAndRelatedWork}

The main application area for \tool{} is in an \gls{HPC} environment.
Computing centers are slowly integrating quantum hardware on-premise~\cite{humble2021}, but they are still isolated entities from a software perspective.
Software development in this area is still piecework and encloses multiple domains studied separately in computer science or physics.
Standardization is progressing slowly; the primary focus is to find a common circuit representation, for example, QIR~\cite{QIRSpec2021}, but runtime components are barely present.
The entire software stack~\cite{schulz2023} is growing increasingly complex, but competitive implementations are still lacking.

A common technique to tackle this challenge is to create a quantum software stack that reimplements successful methods from existing classical counterparts.
The most prominent example is Qiskit~\cite{Qiskit}, offering a full quantum stack; other tools like TKet~\cite{tket2020} and BSQKIT~\cite{younis2021} encapsulate specific functions.
As pointed out, these tools are not yet geared toward \gls{HPCQC}.
A promising approach is to treat the quantum hardware as accelerating \glspl{QPU}, akin to \glspl{GPU}.
XACC~\cite{mccaskey2020} and CUDA Quantum~\cite{cuda2024} are two such frameworks treating \glspl{QPU} as accelerators.
All these tools share the concept of a single quantum backend to which they can offload quantum circuits.  

Using circuit cutting to distribute quantum workloads has been discussed before~\cite{furutanpey2023}, but mainly to reduce errors~\cite{basu2022, bhoumik2023}.
Qurzon~\cite{Chatterjee2022} includes a priority-based, round-robin scheduling algorithm applied to subcircuits resulting from circuit cutting.
This is limited to superconducting devices and optimizes toward noise.
\emph{MILQ}~\cite{seitz2023} is, to our knowledge, still the only scheduler aimed toward the use with heterogeneous quantum hardware.
This includes modalities with unique hardware characteristics, distribution of circuits over independent \glspl{QPU}, and sharing one \gls{QPU} between multiple circuits.

\subsection{Scheduling}\label{ssec:Scheduling}

Scheduling, in its most general form, tries to assign several tasks $J$ over a set of available machines $M$.
Each machine can have different characteristics: capacity, resource restriction, or processing speed.
Similarly, the tasks can differ by their requirements and have complex interdependencies.
Scheduling can be viewed from two perspectives.

As an optimization problem, scheduling tries to find a valid job assignment to machines, respecting all the requirements.
Common examples include job shop, open shop, and flow shop scheduling.
These problems are typically formulated as mathematical programs for which an exact solution is NP-hard to find~\cite{Lenstra1990}.
So-called metaheuristics, including genetic algorithms, simulated annealing, and tabu search, can provide good approximations to optimization problems~\cite{rosenbauer2020}.
They typically do not solve the program directly; instead, they use problem-specific domain knowledge to search and evaluate solution candidates.
The typical quantities of interest include makespan, tardiness, or lateness.

In computing centers, scheduling is a vital component of the runtime environment.
The main concern is to regulate access to the limited computing resources and share them with all users.
Contrary to the optimization problem, the goal is a \emph{fair} solution in a short amount of time.
This includes providing users with their requested hardware and prioritizing urgent jobs while allowing everyone equal access.
SLURM~\cite{Yoo2003} is a widely adopted scheduler, including backfilling and gang scheduling techniques.
Most methods depend on preemption, allowing them to stop and resume jobs during their lifetime.

\subsection{Circuit Cutting}\label{ssec:CircuitKnitting}

Circuit cutting~\cite{Hofmann2009} is a method to decompose a quantum circuit into partitions, each containing fewer qubits than the original circuit. 
Additionally, we require that the fragments form valid quantum circuits that can be executed on quantum hardware. 
Formally, a quantum channel $\mathcal{W}$ describing a gate or a qubit wire is expressed as a sum
\begin{equation}
\label{circuit_cutting_decomposition}
    \mathcal{W}=\sum_{i=1}^L a_i \mathcal{F}_i
\end{equation}
over $L$ local operations $\mathcal{F}_i$ (local with respect to the partitions) where the coefficients $a_i$ are real but possibly negative.
\emph{Gate cutting}~\cite{Hofmann2009, Mitarai2021a,Mitarai2021b, Piveteau2022, Ufrecht2023a, Ufrecht2023b, Schmitt2024,Harrow2024} is the decomposition of a non-local gate or a collection of gates into local unitaries and measurement operations which effectively corresponds to cutting the gate.
If $\mathcal{W}$ comprises all gates connecting two partitions, they become independent through cutting, and the fragmented circuit can be run on hardware with a reduced number of qubits.
Another circuit cutting method is known as \emph{wire cutting}~\cite{Peng2020,Lowe2022, Uchehara2022, Pednault2023,Harada2023,Brenner2023} where one or several qubit wires are effectively cut by decomposition of the identity channel, mathematically describing the wires, into measure and prepare channels. 
To evaluate a circuit cutting protocol as in \cref{circuit_cutting_decomposition} experimentally, at each shot $\mathcal{W}$ is replaced by $\mathcal{F}_i$ with probability proportional to $|a_i|$.
This procedure is referred to as quasi-probability sampling~\cite{Pashayan2015} since the sign of $a_i$ plays a crucial role in determining the expectation value of an observable with respect to the cut circuit.
Note that evaluation is much more efficient when the number of times each channel has to be applied is determined by sampling prior to the experiment. 
This evaluation in batches can be done as long as the circuit-cutting protocol involves at most one-way classical communication~\cite{Piveteau2022, Ufrecht2023b}.

The scheduler described in this work is based on the gate-cutting paradigm.
When optimizing the total runtime for a given set of jobs, the scheduler could strategically cut circuits to maximize the utilization of available qubits. 
However, there is a significant trade-off involved in this process. 
Cutting a circuit substantially impacts the evaluation time of the corresponding job for two main reasons: 
Firstly, all circuit-cutting methods incur a sampling overhead, which is the factor of additional samples required to estimate an expectation value in the cut scenario.
This overhead is quantified  by $\kappa^2$ where $\kappa=\sum_{i=1}^L|a_i|$ is the one norm of the vector of decomposition coefficients.
Secondly, cutting circuits generates $L$ fragments, each differing from another at the position where $\mathcal{F}_i$ was substituted.
The values of $L$ and $\kappa$ increase exponentially with the number of gates cut.
For example, cutting a single CX gate instance leads to $\kappa=3$ and $L=6$~\cite{Mitarai2021a}. 
Consequently, the sampling overhead introduces an exponential increase in the evaluation time for the job, assuming a constant quantum runtime for each shot.
Conversely, suppose all $L$ sub-circuits need to be compiled individually. 
In that case, the exponential increase of compilation time can become the dominant factor~\cite{Harada2023}, as the compilation time for a single circuit is typically much greater than the single shot quantum runtime. 
The scheduler must consider all these aspects and, ideally, within a joint optimization loop, would automatically determine whether or not to cut and the optimal locations for each cut.

\section{Proposal}\label{sec:Proposal}

\tool{} is not intended as a standalone tool.
It is positioned in a stack of multiple interacting tools.
In the Munich Quantum Software Stack~\cite{schulz2023}, \tool{} would take the role of the \emph{resource manager}.
In particular, we assume the existence of pre- and postprocessing steps that the scheduler does not need to deal with.
For example, high-level circuit optimization should happen before submitting a job.
Device-specific optimization, such as mapping and routing, is only practical after a device is specified.

\begin{figure}
    \centering
    \includegraphics[scale=.5]{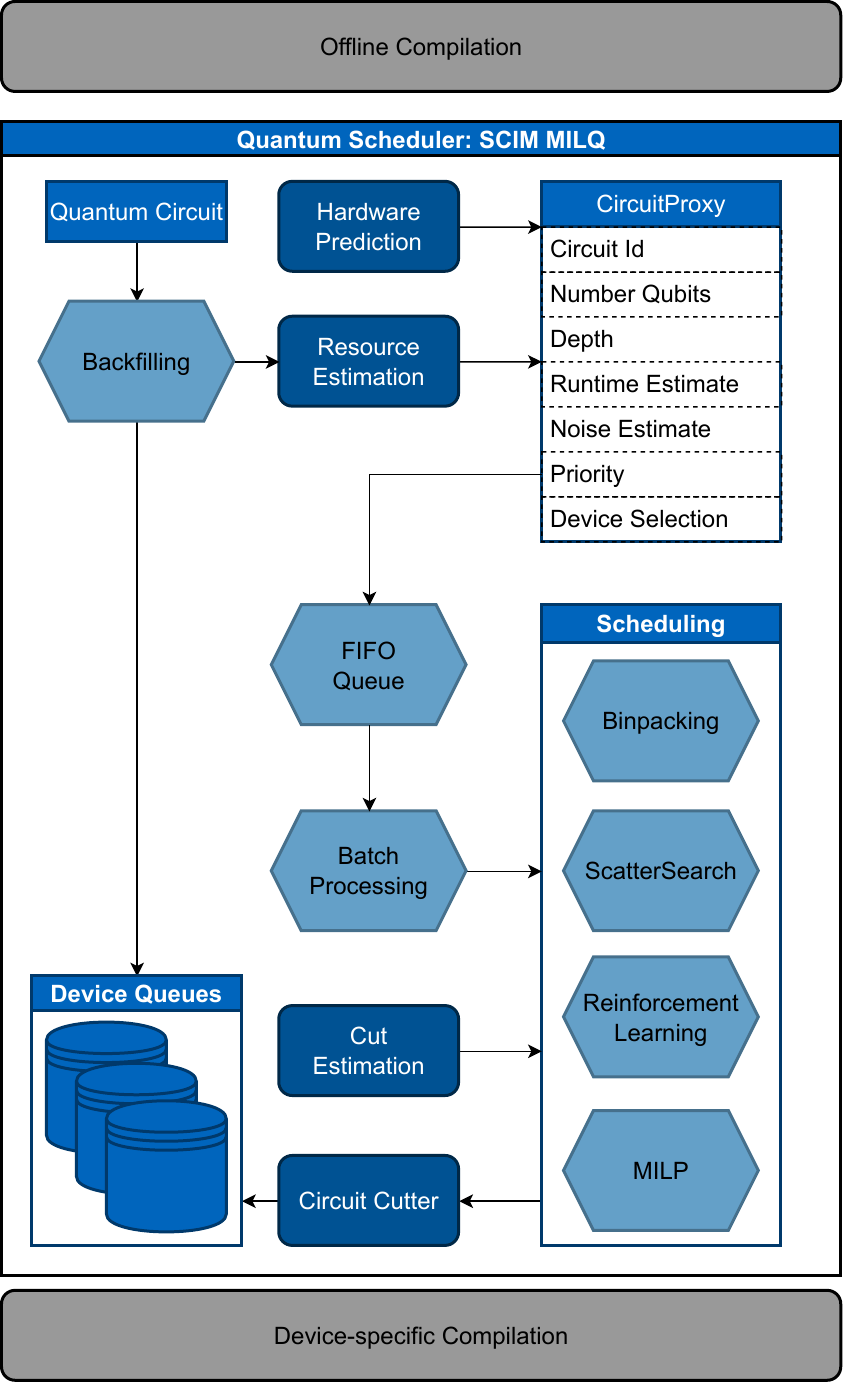}
    \caption{Overview of the intended workflow in \tool.
    The regular scheduling components are drawn as hexagons; the quantum components are rounded rectangles.
    Before job submission, other tools can optimize quantum circuits.
    After the hardware is selected, the device-specific compilation can start.}
    \label{fig:overview}
\end{figure}

Circuit cutting is usually part of the initial circuit generation procedure, which can already include device information.
A typical scheduler is unaware of the choices made during this process, simply scheduling the resulting circuits.
This can lead to suboptimal decisions when hardware selection does not align with the assumptions during cutting.
Inverting the order of these steps is also not sufficient since hardware assignment might be inconsistent after cutting.
Crucially, circuit generation and hardware selection interplay is a significant challenge for a scheduler. 
Instead of considering generation and scheduling separately, we include parts of \emph{both} into a single component.
\Cref{fig:overview} shows how \tool{} processes a quantum circuit. 
The individual components are explained in the following section.
Notably, \tool{} uses proxy objects, rather than the actual circuits, to speed up the computation.

\subsection{Workflow}

\tool{} unifies the quantum components with classical scheduling techniques.
In the \gls{HPC} context, it is common for the user to define a complete algorithm and specify which resources to use.
This would mean that, for traditional schedulers, a user would submit a full circuit as a job.
In our case, however, the program does not need to declare a circuit ab inito, but at least during its runtime.
One goal of \tool{} is to support both extremes, where users want complete control over the hardware or none.
An underlying assumption is that the compute center can manage the quantum hardware freely.
This requires skipping any built-in scheduling mechanisms, typical for cloud-based access.
For simplicity, we also assume a standard interface over all devices, in which every device has a single \gls{FIFO} \emph{device queue}.

As soon as the program constructs the circuit, it can be submitted to the schedule.
Before the circuit is enqueued, it can skip further processing in two scenarios.
Both are only possible if the user does not specify an exact hardware backend to run on.
The related job can immediately start processing if any sufficiently large quantum device is idle.
In the case of smaller circuits, it is also possible to skip the scheduling mechanism if a \emph{device queue} has a timeslot with enough free space to accommodate the job.
This mechanism is referred to as \emph{backfilling} and is commonly found in modern schedulers such as SLURM~\cite{Yoo2003}.

\begin{table}[]
    \centering
    \caption{Parameters of interest for a proxy.}
    \label{tab:proxy}
    \begin{tabular}{ll}
        \toprule
        \textbf{Parameter} & \textbf{Description} \\ \midrule
        \multicolumn{2}{c}{User Input Parameters} \\ \midrule
        $\tau_i$ & The machine preference of job $i$ \\
        $\sigma_i$ & Strictness of the machine preference, $0<\sigma_i$\\
        $\rho_i$ & Priority of job $i$, $\rho_i \in [1, 2, \ldots, 20]$ \\ \midrule
        \multicolumn{2}{c}{Circuit Parameters} \\ \midrule
        $Id_i$ & UUID of the original circuit job $i$ belongs to \\
        $q_i$ & Number of qubits of job $i$ \\
        $d_i$ & Circut depth of job $i$ \\ 
        $b_i$ & (Estimated) Start time of job $i$\\
        $c_i$ & (Estimated) Completion time of job $i$ \\ \midrule
        \multicolumn{2}{c}{At-Runtime System Queries} \\ \midrule
        $\ell(m)$ & Current queue length of machine $m$ \\
        $p(i,m)$ & (Estimated) Processing time of job $i$ on $m$  \\
        $s(i,j,m)$ & (Estimated) Set-up time between job $i$ and $j$ on $m$  \\ 
        $f(i,m)$ & (Estimated) Noise of job $i$ on $m$ \\\bottomrule
    \end{tabular}
\end{table}

Instead of processing the potentially large quantum circuit, \tool{} operates on so-called \emph{proxy} objects.
A summary of the parameters stored by the \emph{proxy} objects is listed in \cref{tab:proxy}.
They refer to their origin by a unique identity and hold all values necessary for the scheduler to operate.
They inherit the information about depth and the number of qubits from the circuit.
Additionally, based on their characteristics, resource estimates are added.
This includes estimates for the processing time and the expected noisiness of the circuit. 
The formulas will be given below.

The \emph{proxy} contains optional user-specified values.
Priority states how urgently a job must be finished.
A higher priority value indicates an earlier deadline.
With a strictness value $\sigma$, the user can also require a specific target hardware.
A high strictness value incentivizes the scheduler to choose the selected hardware, considering this might increase the queueing time.
An initial guess for the best hardware can be made if the user specifies no hardware.
We use the MQT-Predictor~\cite{quetschlich2023} for hardware prediction.
This guess is used as the initial job placement, which in turn is used to seed the heuristics.

After converting the circuit, it is enqueued in a simple \emph{\gls{FIFO} Queue}.
The \emph{scheduling} itself still operates on batches of proxies to reduce complexity.
To realize this, the front of the queue is monitored by a \emph{batch processing} component.
A batch contains the first $n$ items to reach the qubit count threshold $t$:
\begin{equation}\label{eq:threshold}
 \sum_i^n q_i\,\leq t.
\end{equation}
The batch is then handed to the scheduling algorithms, assigning each job $i$ to a machine $m_i\in M$ of the available machines $M$.
Currently, \tool{} supports the following methods:
\begin{enumerate}
    \item \emph{Binpacking}: first-fit decreasing bin packing, only considering the makespan, as implemented in \emph{MILQ}.
    \item \emph{MILP}: Mixed integer linear programming based on the \emph{simple} and \emph{extended} formulation from \emph{MILQ}.
        Both only support makespan optimization and operate on the exact values for $p(i,m)$ and $s(i,j,m)$.
    \item \emph{Scatter Search}: A meta-heuristic approach for optimization; details are described in \cref{ssec:scatter_search}. Supports makespan optimization, including cutting decisions.
    \item \emph{Reinforcement Learning}: A machine learning agent; details are described in \cref{ssec:rl}. The agent includes a joint optimization of noise and makespan, also including cutting decisions.
\end{enumerate}
The latter two methods include circuit-cutting capabilities.
Since constructing the exact subcircuits is expensive, we also use \emph{proxies} in this situation.
For execution, the essential factor is the resulting sampling overhead, as argued below. 
Given the scheduling instructions, the \emph{cut estimation} searches for a low-cost cut.
For example, the scheduler can distribute a circuit of size ten over two devices with five qubits each.
The \emph{cut estimation} then tries to find a partition minimizing the overhead and returns only the number of resulting circuits and their required shots.
New circuits are placed at the end of the schedule, and the user parameters are inherited from their parent.
For each of the resulting subcircuits, we reestimate the runtime (and expected noise in the case of reinforcement learning) based on the approximations detailed below in \cref{ssec:resource_estimates}.
Additionally, we store the information of the parent circuit, which we use to calculate the setup times.
We assume that setting up between similar circuits, resulting from circuit cutting could be significantly faster than a complete reloading procedure.
The scheduler uses resource estimates to approximate the potential start and completion times to evaluate candidate solutions.

As the result of the scheduling phase, a schedule of \emph{proxies} is returned.
The \emph{circuit cutter} component retrieves and cuts the original circuits accordingly.
Hence, the final circuit generation is only completed once the hardware is determined.
Circuits that are scheduled concurrently are combined into a single circuit and submitted to the correct device queue.
For monitoring purposes, the accurate makespan of the circuits can be calculated using the available device information.

While the circuits wait in the queue, other systems can complete the optimization.
We deliberately did not include hardware-specific optimizations, such as mapping and routing in \tool{}\@.
This would dramatically increase the complexity of scheduling, and the potential gain is not apparent.
After execution, the final step is the reconstruction of the measurements.
Using the circuit identifiers, \tool{} can faithfully restore the data.
This should be a separate component to avoid monolithic architecture in the future. 

\subsection{Resource Estimates} \label{ssec:resource_estimates}
Repeatedly recalculating circuit processing time estimates after cutting was a bottleneck for our previous implementation.
This is because we used a complete compilation pass to create a concrete gate schedule and retrieve the exact processing time.
In \tool{}, we solve this in a two-step approach.
First, we use the Azure Resource Estimator~\cite{vandam2023} to estimate the processing time for all circuits once with a good approximation ratio.
During scheduling, whenever we cut a circuit, we approximate the processing times of the proxies based on the original, full processing time and the ratio of circuit depths for the resulting subcircuits $j$ of the original circuit $i$:
\begin{equation}
    p_j = p_i \times \frac{d_j}{d_i}\,.
    \label{eq:processing_time}
\end{equation}
Cutting does not necessarily decrease the depth of a circuit, but some remaining subcircuits might be shorter overall.
Although the estimate from the resource estimator is targeted toward the error-corrected regime, we can use it as long as the circuits are still comparable.
Still, in the future, we want to use device-specific estimates, including potential routing overhead,  to improve the quality of the solution.
As the execution information highly varies between hardware and is not readily available, we choose a generic approach.

Similarly, we estimate the expected noise placing a circuit on a device.
We use Mapomatic~\cite{nation2023} to evaluate the expected noise.
Compared to the processing time, this is device-specific and does not yield a result for circuits larger than the device.
To solve this, we first extrapolate the overall noise by calculating the noise of some subcircuits.
We take the \emph{maximum} noise over all devices for the initial estimate.
The expected noise of a proxy is estimated as
\begin{equation}
    f(j,m)=\max_{m\in M} f(i,m) \times \frac{d_j}{d_i}\,.
    \label{eq:noise}
\end{equation}

Mapomatic estimates the noise based on the coupling graph and the hardware calibration data.
It considers the graph isomorphisms between the circuit connectivity and coupling graphs.
Each possible mapping is associated with a noise estimate.
Typically, it is used to find an optimal mapping, but it can also be used for scheduling~\cite{bhoumik2023}.
We restrict the search depth as we are only interested in an approximation.

As a proxy for the unknown true cost, \emph{cut estimation} determines the sampling overhead and number of jobs resulting from cutting. 
Due to its exponential growth in the number of cut gates, the sampling overhead is expected to be the critical factor in determining whether circuit cutting should be used for a given circuit instance. 
Indeed, cutting one additional CX gate increases the sampling overhead by a factor of nine.
Hence, the inclusion of \emph{cut estimation} into the complete optimization would result in a strong tendency towards solutions with minimal sampling overhead.
As a reasonable approximation, \emph{cut estimation} finds cutting options corresponding to different partition sizes, each with low or near minimal sampling overhead.
Currently, this routine is implemented using brute-force search but will be replaced by an efficient graph-partitioning algorithm in a future version.
Furthermore, the only two-qubit gate currently supported is the CX gate.

As described in \cref{ssec:CircuitKnitting}, it has been argued~\cite{Harada2023} that the overhead from compiling the $L$ circuits, again exponential in the number of gates, can be dominant. 
To sidestep this issue, we assume a single compilation of the original circuit with the gate to be cut as a placeholder. 
At execution time, the channel $\mathcal{F}_i$ is substituted at this circuit position. 
In case of joint gate cutting with reduced sampling overhead~\cite{Piveteau2022, Ufrecht2023b}, $\mathcal{F}_i$ contains two-qubit gates that can be routed efficiently to the topologically constrained hardware using CX-gate resynthesis~\cite{Nash2020}.
While this might slightly increase the depth of the circuits, it seems reasonable to assume that state-of-the-art compilation pipelines can be amended so that this step can be performed in negligible time, i.e., $s(i,j,m)\sim0$.

In summary, the interplay between \emph{cut estimation} and other scheduling components is modeled as follows: Given the sampling overhead, the total number of required shots is calculated. 
The result is distributed over the $L$ circuits. 
The jobs are then passed into the scheduling pipeline, neglecting the extra circuits' compilation latency.

\section{Circuit Scheduling Heuristics}\label{sec:Heuristics}

\begin{algorithm*}[t!]

\SetKwInOut{Input}{Input}
\SetKwInOut{Output}{Output}
\SetKwFunction{init}{InitializePopulation}
\SetKwFunction{new}{GenerateNewSolutions}
\SetKwFunction{improve}{ImproveSolutions}
\SetKwFunction{elite}{SelectTopN}
\SetKwFunction{diverse}{SelectMostDiverse}

\Input{$J:= \text{set of jobs}, M:= \text{list of machines}, N:= \text{number of iterations}$}
\Output{Approximate Schedule}
$population \gets \init{J, M}$\;
$bestSolution \gets \elite{population, 1}$\;
\For{$1..N$}{
    $newSolutions \gets \new{population}$\;
    $improvedSolutions \gets \improve{population}$\;
    $population \gets population \cup newSolutions \cup improvedSolutions$\;
    $eliteSolutions \gets \elite{population}$\;
    $diverseSolutions \gets \diverse{population, topN}$\;
    $population \gets eliteSolutions \cup diverseSolutions$\;
    $currentBestSolution \gets \elite{population, 1}$\;
    \If{$currentBestSolution < bestSolution$}{
        $bestSolution \gets currentBestSolution$\;
    }
}
\KwRet{$bestSolution$}\;
\caption{Scatter search for scheduling}\label{alg:scatter}
\end{algorithm*}

A significant drawback of the linear programming approach used in \emph{MILQ} is the performance-to-result ratio.
Solving the program improves the schedule but at a high computing cost.
Introducing domain-specific knowledge, scheduling can be solved with so-called meta-heuristics with good approximation ratios~\cite{Kalra2021}.
Typically, these heuristics include mechanisms to search for local improvements while exploring most of the search space.
We use \emph{Scatter Search}, which has been applied to scheduling problems previously~\cite{Kalra2021}.
While this is already an improvement over the existing scheduling, not all problems can be addressed sufficiently.
The performance still suffers from exponentially large search spaces, and simultaneous multi-objective optimization for noise and makespan is almost impossible.
To provide these capabilities when necessary, \tool{} offers a second mode of operation.
We trained a reinforcement learning agent for a restricted set of scenarios. 
It optimizes for a mix of noise and makespan and can freely choose to cut any circuit.

Both approaches are based on a preliminary schedule, which they evaluate and update regularly.
This schedule assigns circuit proxies to timeslots, roughly grouping circuits with similar starting times. 
A subroutine reconstructs the total running time to evaluate a schedule.
First, we iteratively calculate the completion times for all circuits based on the last completed circuit and the setup time between them:
\begin{align}
    c_0 &= p(0,m)\,, \\
    c_i &= p(i,m_i) + c_{i-1} + s(i-1,i,m_i)\,.
\end{align}

Then, we score each machine $m$ considering priority values, hardware suggestions, and current queue lengths for all available machines by
\begin{equation}\label{eq:score}
    \mathbf{P_m} =  \ell(m) + \max_i( c_i \times \rho_i \alpha + \delta_{i,\tau_i} \times \sigma_i\beta)\,. 
\end{equation}
Using the Kronecker-Delta $\delta_{i,\tau_i}$, which indicates circuit $i$ is scheduled on its suggested machine $\tau_i$ and the user-specified strictness value.
If the user did not specify the hardware, we set $\sigma_i=0$.
The system can be fine-tuned by specifying the values $\alpha$ and $\beta$, which control user-defined parameters' importance.
We generally choose priority as the primary parameter, but this is up to the system provider.

The total cost $\mathbf{P_\text{max}}$ of a schedule then is 
\begin{equation}\label{eq:cost}
    \mathbf{P_\text{max}} = \max_{m\in M} \mathbf{P_m}\,.
\end{equation}
The goal of the scheduling algorithms below is to minimize this value.
After the schedule is constructed, the circuit cutter replaces the proxies with their actual counterparts.

\subsection{Scatter Search}\label{ssec:scatter_search}

The scatter search procedure is summarized in \cref{alg:scatter}.
The algorithm generally tries to iteratively improve its set of candidate solutions and continuously introduce new unexplored candidates.
To keep the size of the search space manageable, only a restricted subset of the population is maintained after each iteration.
A primary benefit of scatter search is its embarrassingly parallel nature.
Our implementation runs entirely in parallel, effectively increasing the number of iterations for the search.
One could also communicate the best solutions after each step, but this would incur synchronization.
\Cref{tab:params} contains the available hyperparameters. 

\begin{table}[]
    \centering
    \caption{Scatter search hyperparameters}
    \label{tab:params}
    \begin{tabular}{ll}
        \toprule
        \textbf{Parameter} & \textbf{Description} \\ \midrule
        $N$ & Number of total iterations \\
        $eliteSolutions$ & The number of elite solutions to keep \\
        $initializations$ & List of initialization procedures \\
        $numSwaps$ & The number of swaps during local search \\
        \bottomrule
    \end{tabular}
\end{table}

\subsubsection{Initialization}

We apply circuit cutting for initialization, generating candidate solutions based on different cut choices.
We provide multiple cutting variants, allowing for a diverse initial population.
For example, we use the original bin-packing-based approach from \emph{MILQ}~\cite{seitz2023}. 
This guarantees that the solution will be at least as good as the baseline. 
A more sophisticated approach is to find suitable partitions for cutting.
We use the insight from \cref{ssec:resource_estimates} to minimize the sampling overhead and the number of circuits.
This is still done greedily but gives better results for complex circuits.
For each cutting variant, a candidate schedule is generated by cutting all circuits in a batch according to the variant.
The resulting proxies are then assigned according to their machine preference.

\subsubsection{Local Search}

The local search is executed in the \verb|GenerateNewSolutions| function at the start of each iteration.
It updates a schedule by applying one of two possible updates to an existing schedule.
It swaps two randomly chosen jobs between timeslots on one machine or between machines.
A dummy job can be selected to move a job instead of swapping.
A configuration parameter controls the number of swaps per round.

\subsubsection{Active Improvement}
\verb|ImproveSolutions| improves the existing population by reducing the schedule proactively.
Jobs from the machine with the longest makespan are moved to the machine with the shortest.
This is achieved by selecting one random timeslot and distributing all its jobs over timeslots on the new machine.
The resulting schedules can be invalid, but this is explicitly allowed.

\subsubsection{Selection}

The population update consists of two parts.
First, the candidates with the lowest cost $\mathbf{P_\text{max}}$ are selected.
Candidates who differ most from the existing population are retained to encourage diversification.
We use a custom distance metric based on the Hamming distance for two schedules, $S_1$ and $S_2$.
The metric compares the position of all jobs in the respective schedule for each machine.
It uses the current machine assignment $I_m=\{\text{circuit }i: i \text{ is assigned to machine } m\}$ and its correpsonding length $\Vert I_m\Vert_\ell=\max_{i\in I_m} c_i$:

\begin{align}
    d(S_1,S_2)&=\sum_{m\in M}\sum_{I_m} d(i_{S_1},i_{S_2})\,,\\
    d(i,j)&=\begin{cases}
        |b_i - b_j| & j \in I_m\\
        \max(\Vert I_m\Vert_\ell,\Vert J_m\Vert_\ell) & \text{otherwise}
    \end{cases}\,.
\end{align}

\subsection{Reinforcement Learning}\label{ssec:rl}

We also implemented a \gls{RL} agent to complement the shortcomings of the previous approach. 
In contrast to meta-heuristics, simultaneous optimization of multiple objectives can be achieved more easily.
We choose a simple linear combination of makespan and expected noise as a reward function $R$ controlled by two configuration parameters $\mu$ and $\nu$ :
\begin{equation}
    R = -(\mu \times \mathbf{P_\text{max}} + \nu \times \sum_m\sum_i f(i,m))\,.
\end{equation}

The action space consists of four actions parametrized by the circuits they act on
\begin{itemize}
    \item Cut: Cut a circuit $i$ at qubit $n$. Invalid cuts are disincentivized with a penalty.
    \item Move: Moves a circuit $i$ from timeslot $t_i$ on machine $m$ to timeslot $t_j$ on machine $n$. 
            The case $n=m$ is allowed, but the case $t_i=t_j$ is not.
    \item Swap: Swaps circuits $i$ and $j$ in the schedule
    \item Terminate: If the schedule is in a valid configuration, the search can be ended early. 
\end{itemize}
This agent is straightforward and depends on an initially feasible schedule; We plan to improve its capabilities in future versions.
To train the agent, we use Proximal Policy Optimization~\cite{schulman2017} as implemented by the \emph{Stable Baselines 3}~\cite{stable-baselines3} Python package.
The training is executed for a total number of $5000$ iterations.
Afterward, we extract the last schedule as an observation from the agent.
This schedule is then forwarded to the circuit cutter to recreate the cuts on the original circuits.

\section{Results}\label{sec:Results}

We evaluate \tool{} by creating hypothetical scenarios in which multiple circuits are submitted to the system.
The configuration parameters are displayed in \cref{tab:config}; additionally, we have disabled backfilling.
We opt for two hardware scenarios based on simulators.
First, we use two devices with five and seven qubits and identical set-up times.
As a second scenario, we add a five-qubit device with higher set-up times.
Each device is prepopulated with a random queue length to simulate load.
The test circuits of various sizes are generated using MQT~Bench~\cite{quetschlich2023_2}.
Each circuit is assigned random values for $\tau_i$, $\sigma_i$ and $\rho_i$ (c.f. \cref{tab:proxy}).
We simulate ten batches for the \emph{baseline}, \emph{heurisitc}, and \emph{reinforcement learning} schedulers.

\begin{table}[t]
    \centering
    \caption{Default scheduling configuration used for the experiments.}
    \label{tab:config}
    \begin{tabular}{cc}
    \toprule
        \textbf{Parameter} & \textbf{Value}  \\ \midrule
         Batch size & 5 \\
         $\alpha$   & 1 \\
         $\beta$    & 1 \\
         $\mu$      & 0.5 \\
         $\nu$      & 5 \\ \bottomrule 
    \end{tabular}
\end{table}

After generating the schedules, we use the \emph{circuit cutter} to create the device queues accordingly.
Using the actual circuits, we update the runtime and noise estimates to calculate the final metrics.
We evaluate the results using three criteria:
\begin{enumerate}
    \item \emph{Makespan}: The maximum termination of the latest job, excluding the existing queue.
    \item \emph{Schedule Cost}: $\mathbf{P_\text{max}}$ as defined in \cref{eq:cost}.
    \item \emph{Noise}: $\mathbf{F}=\sum_m\sum_i f(i,m)$.
\end{enumerate}

\begin{figure}[t]
    \centering
    \includegraphics[scale=0.6]{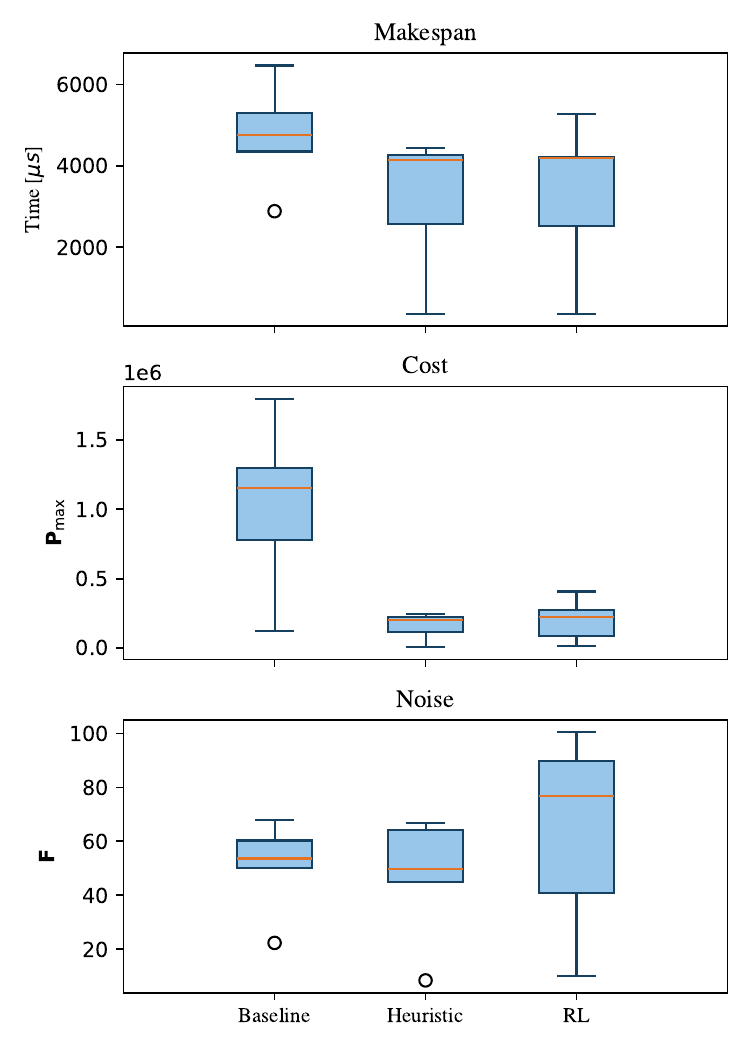}
    \caption{Results for the benchmark simulation across the makespan, schedule cost $\mathbf{P_\text{max}}$, and noise $\mathbf{F}$; lower is better. 
    We illustrate the setting of three devices with sizes five, five, and seven. 
    The results for the second setting are very similar.}
    \label{fig:results}
\end{figure}

\Cref{fig:results} reports the results for the setting $(5,5,7)$. 
We see an improvement in our $\mathbf{P_\text{max}}$ cost of about \SI{80}{\percent}.
As expected, the heuristic outperforms the baseline algorithm, as the baseline does not include the user-defined parameters.
The reduced cost of choosing better cost is also reflected in the makespan, where the heuristic improves \SI{25}{\percent} on average.
The total noise is similar for both the baseline and the heuristic, with a difference of only \SI{2}{\percent} in favor of the baseline.
In favorable settings, the \gls{RL} approach can improve the noise by \SI{10}{\percent}, but overall, its results are highly dependent on the initial schedule.
Performance-wise, all three approaches perform similarly.
The \emph{circuit cutter} to create the final circuits is the bottleneck.

\section{Future Work} \label{sec:FutureWork}

With \tool{}, we have made considerable progress toward a production-ready scheduler.
But we have tested it only in an isolated environment, without heavy load.
The next step is integration into the Munich Quantum Software Stack~\cite{schulz2023} hosted by the Leibniz Supercomputing Center.
This will offer a multifaceted production environment, including multiple hardware modalities and access patterns.
Further performance optimization is necessary to offer a competitive product.
This includes fine-tuning the scheduling algorithms to the use cases and the general execution efficiency.
The hyperparameters for the scheduling heuristics especially need to be analyzed thoroughly.
Once this is completed, advanced features like gang scheduling~\cite{Yoo2003} and co-scheduling quantum and classical hardware can be included.

The quantum components also have room for improvement.
Implementing advanced circuit cutting techniques in the circuit cutter, especially joint cutting, can significantly improve the results.
This could even go so far as to include precompilation of circuits toward improving the ``cuttability'' of circuits. 
Noise reduction methods could also be integrated into the workflow.
For example, one could include simulators to perform noise-free simulations for some subcircuits when necessary.
Further, we want to incorporate more hardware, even analog quantum devices, which might require a hierarchical approach to handle the complexity.

So far, we have based \tool{} on resource estimates for general hardware configurations.
The scheduling algorithms work best with accurate input.
This requires thorough benchmarking of available hardware regarding noise, processing and setup times, and evaluation of reconstruction from different modalities.
We hope that in the future, switching between subcircuits belonging to the same partition will be significantly faster compared to the current state.
Once quantum memory becomes a reality, preemption in the scheduling will be possible.
This opens up many opportunities to further improve \tool.

\section{Conclusion} \label{sec:Conclusion}

\tool{} represents an essential ingredient for integrating quantum computing with \gls{HPC}.
It combines multiple quantum and classical concepts and utilizes them to enable a typical \gls{HPC} workflow.
It is highly customizable and offers multiple modes of operation, including a novel joint optimization for makespan and noise.
We show its viability in realistic scenarios while delivering on-par performance compared to previous work.
Further, we introduce a new evaluation metric incorporating classical scheduling concepts.
Using this metric, we achieve overall improvements of \SI{80}{\percent} compared to existing work, while simultaneously improving makespan by \SI{25}{\percent}.

\section*{Acknowledgment}
The research is part of the Munich Quantum Valley~(MQV), which is supported by the Bavarian state government with funds from the Hightech Agenda Bayern Plus. Moreover, this project is also supported by the Federal Ministry for Economic Affairs and Climate Action on the basis of a decision by the German Bundestag through project QuaST, as well as by the Bavarian Ministry of Economic Affairs, Regional Development and Energy with funds from the Hightech Agenda Bayern. Further funding comes from the German Federal Ministry of Education and Research~(BMBF) through the MUNIQC-SC project.

\IEEEtriggeratref{29}
\bibliographystyle{IEEEtran}
\bibliography{references}
\end{document}

%% file: glosfile.tex
\newacronym{MILP}{MILP}{mixed-integer linear program}
\newacronym{AWG}{AWG}{arbitrary waveform generator}
\newacronym{NISQ}{NISQ}{noisy intermediate-scale quantum}

\newacronym{CPU}{CPU}{central processing unit}
\newacronym{GPU}{GPU}{graphics processing unit}
\newacronym{QPU}{QPU}{quantum processing unit}
\newacronym{HPC}{HPC}{high performance computing}
\newacronym{LRZ}{LRZ}{Leibniz Supercomputing Centre}
\newacronym{DSL}{DSL}{domain-specific language}
\newacronym{eDSL}{eDSL}{embedded domain-specific language}
\newacronym{MPI}{MPI}{Message Passing Interface}
\newacronym{QRAM}{QRAM}{quantum random access machine}
\newacronym{HQCC}{HQCC}{heterogeneous quantum-classical computation}
\newacronym{API}{API}{application programming interface}
\newacronym{QPT}{QPT}{quantum programming tool}
\newacronym{QC}{QC}{quantum computing}
\newacronym{ISA}{ISA}{instruction set architecture}
\newacronym{RISC}{RISC}{reduced instruction set computer}
\newacronym{RAM}{RAM}{random access memory}
\newacronym{IR}{IR}{intermediate representation}
\newacronym{SQRAM}{SQRAM}{sequential QRAM}
\newacronym{HPCQC}{HPCQC}{high performance computing - quantum computing}
\newacronym{QIR}{QIR}{Quantum Intermediate Representation}
\newacronym{SQIR}{SQIR}{standardized quantum intermediate representation}
\newacronym{AOT}{AOT}{ahead-of-time}
\newacronym{JIT}{JIT}{just-in-time}
\newacronym{FIFO}{FIFO}{first-in-first-out}
\newacronym{RL}{RL}{reinforcement learning}